\documentclass[conference]{IEEEtran}
\IEEEoverridecommandlockouts

\usepackage{cite}
\usepackage{amsmath,amsfonts}
\usepackage{mathtools}

\usepackage{graphicx}
\usepackage{graphics}
\usepackage{textcomp}
\usepackage{xcolor}
\usepackage{booktabs}
\usepackage{tikz}
\usepackage{bm}
\usepackage[export]{adjustbox}
\usepackage{comment}
\usepackage{enumitem}
\usepackage[normalem]{ulem}
\usepackage{pbox}
\usepackage{multirow}
\usepackage{makecell}
\usepackage{algorithm}
\usepackage{algpseudocode}
\usepackage{array}
\usepackage{dblfloatfix}
\usepackage{hyperref}

\begin{document}

\title{$\Delta$-Motif: Parallel Subgraph Isomorphism via Tabular Operations for Scalable Layout Selection}

\author{
\IEEEauthorblockN{Yulun Wang\textsuperscript{1},
Esteban Ginez\textsuperscript{1},
Jamie Friel\textsuperscript{2},
Yuval Baum\textsuperscript{1},
Jin-Sung Kim\textsuperscript{3},
Alex Shih\textsuperscript{1},
Oded Green\textsuperscript{3}}
\IEEEauthorblockA{\textsuperscript{1}Q-CTRL Inc., Los Angeles, USA\\
\textsuperscript{2}Oxford Quantum Circuits, United Kingdom\\
\textsuperscript{3}NVIDIA, USA}
}

\maketitle
\pagestyle{plain}
\thispagestyle{plain}

\begin{abstract}
    As quantum processors scale to hundreds and eventually thousands of qubits, selecting optimal qubit layout during circuit compilation becomes a critical bottleneck. Layout selection requires solving subgraph isomorphism, a fundamental graph analysis problem: enumerating all occurrences of a pattern graph within a larger data graph while preserving structural relationships. This NP-complete problem underpins a wide range of applications beyond quantum compilation, including biological network analysis and data mining. Classical algorithms such as VF2 rely on backtracking search, which suffers from sequential bottlenecks that limit scalability on modern parallel hardware.
    
    We introduce $\Delta$-\textit{Motif}, a GPU-accelerated subgraph isomorphism algorithm that reformulates the problem through the lens of database operations. Our key insight is to represent graphs in tabular format, transforming subgraph isomorphism into database primitives including joins, sorts, merges, and filters. $\Delta$-\textit{Motif} decomposes graphs into small building blocks called motifs and systematically combines these using scalable relational operations. By leveraging highly optimized libraries from the NVIDIA RAPIDS ecosystem, our solution achieves massive parallelization without specialized low-level GPU kernels, while remaining portable across any system supporting standard relational primitives.

    We evaluate $\Delta$-\textit{Motif} on graph workloads motivated by quantum computing, which capture the structural properties of quantum circuits and device topologies arising in layout selection. On scaling benchmarks with large device topologies, $\Delta$-\textit{Motif} substantially outperforms established methods such as VF2 and GSI, achieving speedups of up to $595\times$ on GPU architectures. On realistic workloads using a state-of-the-art 156-qubit quantum processor, our end-to-end transpilation benchmarks show up to two orders of magnitude faster layout generation.
\end{abstract}

\begin{IEEEkeywords}
Subgraph Isomorphism, GPU Acceleration, Quantum Circuit Compilation, Parallel Computing, Graph Algorithms
\end{IEEEkeywords}

\section{Introduction}
\label{sec:intro}
Graphs have emerged as one of the most flexible and powerful data structures to represent complex relationships, leading to their ubiquitous presence across various computational domains. From modeling molecular interactions in bioinformatics \cite{BioinformaticsAndGraphs} and tracking social connections in network analysis \cite{SocialNetworksAndGraphs} to mapping dependencies in cybersecurity systems \cite{CybersecAndGraphs}, graphs provide an intuitive and mathematically rigorous framework for capturing the intricate patterns that define real-world systems. 

Central to many graph-based applications is the subgraph isomorphism problem, which seeks to find all instances of a pattern graph within a larger data graph while preserving structural relationships. This problem is known to be NP-complete \cite{Cook1971}, yet it remains essential for tasks including circuit mapping in quantum devices \cite{Nation2023}, motif discovery in biological networks \cite{BiochemAndSubgraphIso}, social networks analysis \cite{Dasgupta2020DiscoveringIS}, data mining \cite{Zhao2012LargeSC} and drug discovery\cite{Aung2008BSAlignAR}. The ubiquity of its application to various use cases can create critical bottlenecks in compute and workload execution. Driving efficiencies for subgraph isomorphism will alleviate key barriers to the incorporation of emerging technologies like quantum computing into high-performance computing environments, and thereby its adoption for meaningful impact.

Traditional approaches such as VF2 \cite{Cordella2004}, VF2++ \cite{Juttner2018}, and VF3 \cite{Carletti2019VF3L} rely on backtracking with a largely DFS-like traversal, making them inherently sequential and hard to parallelize. Code specialization techniques have been proposed to mitigate these limitations \cite{Dryadic2021}, and a range of systems have explored distributed execution, accelerator-based designs, and commercial data-processing frameworks such as Hadoop \cite{Zhang2022}, Cassandra \cite{HUGE2021}, and Spark \cite{Wang2021}. Workloads also vary widely---weighted or unweighted graphs, small structured or large random patterns, and diverse connectivity---so no single algorithm is optimal across all settings, and existing work spans many complementary approaches. We discuss these trade-offs in more detail in Sec. \ref{sec:background}.

A key motivation for this work is the use of subgraph isomorphism in quantum circuit compilation, where the pattern graph represents a circuit's qubit interaction structure and the data graph represents the physical connectivity of the target quantum processor. Recent work by Nation et al. \cite{Nation2023} applied a Rust-based implementation of VF2 \cite{Rustworkx2022} in this setting, illustrating both the practicality and limitations of classical backtracking approaches. In our own evaluation, we found that such algorithms limit effective utilization of hundreds of CPU cores and modern GPU architectures.
In quantum computing, the pattern graphs of interest are relatively large, typically dozens to around one hundred vertices, and at these sizes, many existing parallel subgraph isomorphism algorithms perform poorly or fail to scale effectively, whereas our algorithm remains efficient.

To overcome the sequential bottlenecks and limited parallelism of existing subgraph isomorphism enumeration, we developed a novel data-science centric solution that reformulates the problem as a series of tabular data operations. By representing both data and pattern graphs in tabular format, subgraph isomorphism is transformed into fundamental database operations such as joins, sorts, merges and filters. The algorithm can be implemented on any database platform supporting standard relational operations, from traditional databases to in-memory frameworks such as Pandas~\cite{mckinney2011pandas}. For GPU acceleration, we build upon the NVIDIA RAPIDS ecosystem\cite{Cupy2017, cudf-25.6.0, Rapids}, which serves as a drop-in replacement for CPU-based libraries and ensures that future library-level improvements directly benefit our algorithm. As a result, the algorithm inherits massive parallelism from existing optimized libraries, delivering strong GPU performance through familiar data science tools, without requiring specialized low-level programming.

\section{Related Work}
\label{sec:background}
The VF2 algorithm \cite{Cordella2004} uses a depth-first backtracking strategy to incrementally construct partial isomorphisms while applying pruning rules to maintain consistency. 
Several variants have addressed its scalability limitations: VF2++ \cite{Juttner2018}, VF3 \cite{Carletti2017}, and VF3L \cite{Carletti2019VF3L} improve search efficiency through more aggressive pruning strategies, while other CPU-oriented work explores code specialization \cite{AutoMine2019,Dryadic2021}, memory reduction, and algorithmic refinements \cite{Carletti2019VF3P,BEE2025,Li2025,Circinus2023,Bi2016}. 
Although these extensions can accelerate the discovery of individual solutions, they do not fundamentally address the limitations of the tree-based search structure. As a result, these approaches offer limited scalability on modern hardware architectures designed to exploit massive parallelism, particularly GPUs.

Recent work has explored GPU acceleration for subgraph isomorphism \cite{FastSubgraphmatchingGPU2015,GSI2020,cuTS2021,DGSM2022,Chen2022,FastGSM2020,GSI2020,GAMMA2023,SGSI2023} through breadth-first expansion, vertex-centric execution, and specialized data structures. However, most rely on custom GPU kernels, hardware-specific optimizations, and extensive low-level tuning that create significant barriers to adoption. 

Complementary approaches target scalability through indexing and multi-phase execution \cite{FG-Index-Graph-Databases,GraphIndexing-Frequent-Structure,BICE2023} or distributed and disk-based models \cite{DUALSIM2016,SEED2016,HUGE2021,Zhang2022,Wang2021}, but typically incur higher runtime overheads or target smaller pattern sizes.


Beyond these system-level approaches, our work also draws ideas from a different subfield in graph processing. The concept of \emph{motifs} is well established in graph analytics to characterize recurrent structures in biological networks\cite{MotifsAndBio} and network classification tasks\cite{MotifsAndClassif}. However, prior work has primarily focused on motif discovery and counting, rather than using motifs as computational building blocks for solving subgraph isomorphism. 
EGSM \cite{EGSM2023} and ESCAPE \cite{ESCAPE2017} decompose matching tasks into smaller components, an idea related in spirit to our approach. However, these methods rely on specialized data structures and search strategies, whereas our algorithm formulates decomposition and recombination entirely through relational operations.Several approaches also employ join-based formulations \cite{DGSM2022,FastGSM2020,Lai2022}, but assume join-only execution or rely on specialized data structures, increasing implementation complexity and reducing portability.






Existing solutions also target a wide range of pattern types, which makes direct comparisons difficult. Many focus on small patterns: EGSM \cite{EGSM2023} reports results for up to 16 vertices, cuTS \cite{cuTS2021} and ESCAPE \cite{ESCAPE2017} evaluate up to 5 vertices, and other work emphasizes small clique enumeration \cite{Chen2022}. Performance is also strongly influenced by data graph properties; for instance, cuTS targets sparse road networks, while SGSI \cite{SGSI2023} operates on weighted graphs with different computational trade-offs.

In contrast, our algorithm represents both intermediate and final results using standard tabular data structures, benefiting directly from ongoing optimizations in data processing libraries. 
By relying on open-source frameworks with well-defined APIs, our approach prioritizes programmability, portability, and long-term maintainability while maintaining strong performance on modern parallel hardware.

\subsection{Subgraph Isomorphism Enumeration in Quantum Circuit Compilation}
\label{subsec:quantum_compilation}
As quantum hardware steadily advances, near-term processors have emerged as promising platforms for exploring algorithmic realizations to demonstrate quantum advantage \cite{Preskill2018, Kim2023, Sachdeva2024}. However, these systems remain fundamentally constrained by imperfect control \cite{Krantz2019, Mundada2023, Dat2021, Muller2019}, hardware variability \cite{Ravi2023, Tannu2019}, and short coherence times \cite{Stilck2021}. Quantum circuit compilation plays a critical role in minimizing operational overhead and mitigating error accumulation to improve execution fidelity \cite{Nation2025}. Given a quantum device and a quantum circuit, the compilation pipeline aims to realize the target algorithm with fewer operations and a shorter execution time, thereby reducing the impact of noise. These procedures collectively define the topology and operation count of the compiled circuit before execution on hardware.

Crucially, not all qubits (the information carriers in quantum computers) in a quantum processor offer the same operational quality~\cite{Krantz2019, Tannu2019}. Due to the inherent fragility of quantum systems and the challenges associated with device calibration, processors often contain faulty or low-fidelity qubits and couplings. As a result, the choice of which subset of physical qubits to utilize for executing a compiled circuit can significantly influence the circuit's overall computational fidelity. For the same circuit, different mappings of circuit logical qubits into hardware physical qubits, referred to as layouts, may yield significantly different execution outcomes \cite{Nation2023, Hartnett2024}. This sensitivity to choice of layout has motivated the development of systematic layout selection methods, typically structured as a two-phase process: layout generation and score-based ranking. Layout generation explores search space of candidate mappings by enumerating subgraph isomorphisms between the circuit's interaction graph and the hardware's connectivity graph, after which valid layouts are evaluated using device-specific quality metrics to select the most promising one. We accelerate this layout generation stage and integrate such fidelity-based scoring directly into $\Delta$-\textit{Motif}, evaluating the complete layout-selection pipeline in Sec.~\ref{results:transpilation}. Machine 
learning approaches have also been explored for ranking candidate layouts \cite{Hartnett2024}, but these operate on precomputed candidates and do not address layout generation, the most computationally demanding stage.


A distinctive characteristic of quantum layout selection is that the pattern graphs (circuits) are relatively large, typically on the order of dozens to around one hundred vertices in state-of-the-art workloads. This contrasts with many traditional subgraph matching workloads that emphasize small pattern graphs (with 10 or fewer vertices) within very large data graphs. While much of the existing methods focus on scaling to larger data graphs, scaling pattern graph size introduces its own challenges, as larger patterns induce substantial search spaces and many valid embeddings. As quantum devices continue to scale, even larger interaction patterns are expected to become relevant, motivating algorithms capable of efficiently handling larger pattern graphs while exploiting the structured connectivity commonly observed in quantum hardware topologies.

Moreover, data graphs representing the physical qubit connectivity in quantum hardware typically remain static over time, meaning a significant portion of the pipeline can be precomputed and cached. This use-case-specific feature enables the reuse of intermediate results across multiple compilation runs, delivering further speedups in practice when precomputation costs are excluded.

\section{Data-Centric Subgraph Isomorphism Algorithm}
\label{sec:algorithm}
In this section, we introduce a data-science-inspired method for the subgraph isomorphism problem. Instead of exhaustively matching full patterns in the data graph via node-by-node traversal, we adopt a modular strategy based on motif decomposition, which we call $\Delta$-\textit{Motif}, reflecting the decomposition of pattern graphs into smaller components. In Section \ref{sec:search_and_prune}, we draw an analogy between $\Delta$-\textit{Motif} and the $\Delta$-\textit{Stepping} algorithm for single-source shortest path (SSSP) problems \cite{meyer2003delta}, which shares a similar trade-off between additional work and improved parallelism. The details of this analogy become clearer after introducing the algorithm and its performance characteristics.

\subsection{Preliminaries}

\paragraph*{\bf Subgraph Isomorphism}\
We formally define a graph as a tuple $G = (V, E)$, where $V$ is a finite set of vertices and $E \subseteq V \times V$ is the set of undirected edges. Let the data graph be denoted as $G_d = (V_d, E_d)$ and the pattern graph as $G_p = (V_p, E_p)$. We say that $G_p$ is isomorphic to a subgraph of $G_d$ if there exists an injective mapping $f : V_p - V_d$ such that $(u, v) \in E_p \iff (f(u), f(v)) \in E_d$. The subgraph isomorphism enumeration problem is to find all such mappings $f$ from $G_p$ into vertex-induced subgraphs of $G_d$ that preserve adjacency as defined above. Throughout this work, we assume all graphs are unlabeled, undirected and connected.

\begin{figure}[t]
    \centering
    \includegraphics[width=0.32\textwidth]{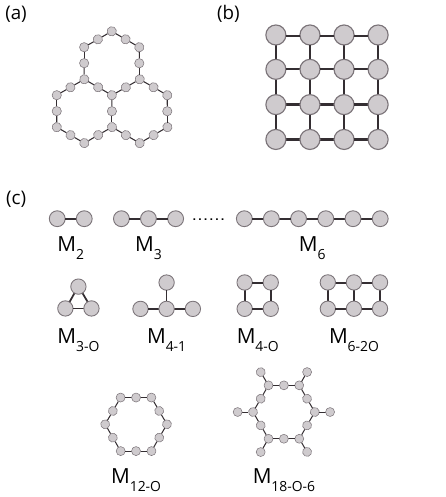}
    \caption{
    (a) Heavy-hex and (b) 2D square grid lattice structures used as data graphs in our benchmark tests. These lattices represent common qubit connectivity topologies in quantum hardware and serve as the input data graphs for quantum layout selection. The corresponding set of representative motifs is shown in (c).
    }
    \label{fig:motifs}
\end{figure}

\paragraph*{\bf Motifs}\
Motifs are size-limited, well-defined subgraphs that capture local structural patterns within a graph. Common examples include linear paths (e.g., $a - b - c$), triangles ($a - b - c - a$), and other frequent occurring configurations such as squares, stars, or short cycles. In this work, we pre-define a set of motifs $\{M_0, M_1, \dots, M_k\}$, where each $M_i = (V_i, E_i)$ is a small, connected subgraph of the data graph. Fig.~\ref{fig:motifs} illustrates the lattice structures used as data graphs in our benchmarks, together with the representative motifs derived from them. We have attempted to keep the naming of the motifs fairly simple: $M_i$ denotes a path of $i$ vertices, while additional suffixes (e.g., “-O” for cycles or “-1” for one branch) indicate common variations.

These motifs serve as the building blocks for decomposing both data and pattern graphs into sequences of motif instances, which can then be incrementally joined to reconstruct larger subgraph matches. As we show in Section~\ref{results:motif_selection}, the choice and ordering of motif set directly influence performance. Selecting motifs requires balancing expressiveness with computational cost, and is informed by the structural characteristics of both the data and pattern graphs.

We denote the set of all embeddings of a motif $M_i$ in the data graph by $Res(M_i)$. Each element in $Res(M_i)$ is a tuple of vertex indices that together form a subgraph isomorphic to $M_i$. These embeddings are stored in tabular form, where each row corresponds to a single match and each column to a vertex position in the motif template. This tabular format allows subsequent processing, such as joining embeddings and enforcing constraints, to be performed using scalable dataframe operations.

\begin{figure*}[h]
    \centering
    \includegraphics[width=0.80\textwidth]{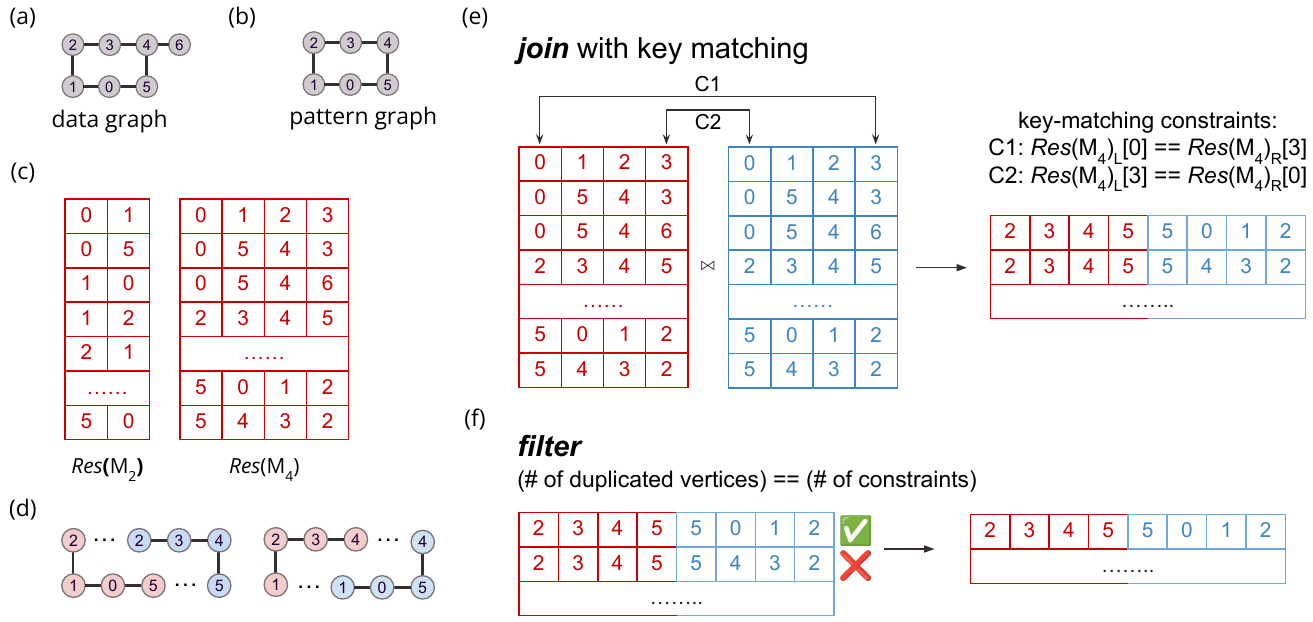}
    \caption{
    A motivating example illustrating subgraph isomorphism: (a) the input data graph and (b) the pattern graph. The decomposition uses two motifs, $M_2$ and $M_4$, along with (c) their corresponding isomorphic embeddings in the data graph, $Res(M_2)$ and $Res(M_4)$. Panel (d) shows two possible decompositions of the pattern graph using $M_4$. Both are valid, but for this example, we use the second decomposition to demonstrate the workflow. Panels (e) and (f) visualize the \textit{join} and \textit{filter} operations in tabular form, which are used to reconstruct valid matches of the pattern graph from motif embeddings. The tables are color-coded to distinguish $Res(M_4)_L$ and $Res(M_4)_R$. The two join key constraints, labeled $C1$ and $C2$, are highlighted with arrows pointing to their corresponding columns.
    }
    \label{fig:motivating_example}
\end{figure*}

\subsection{The $\Delta$-Motif Approach}
\label{subsec:motif_method}
To motivate our approach, we begin with a simple example. Consider a data graph $G_d$ with 7 nodes, shown in Fig. \ref{fig:motivating_example}(a), and a pattern graph $G_p$ defined as a 6-node undirected cycle shown in Fig. \ref{fig:motivating_example}(b). Rather than trying to match the entire cycle directly in the $G_d$, we firstly decompose $G_p$ into a set of smaller, overlapping motifs, $\mathcal{M}=\{M_0, \dots, M_i\}$. Each motif captures a localized structure that can be matched independently. For each $M_i$, we compute the set of all isomorphic embeddings in the data graph, denoted $Res(M_i)$. These partial embeddings are then incrementally combined through \textit{join} operations, followed by a \textit{filter} step to ensure that only valid matches of the full pattern are retained. The final result, $Res(G_p)$, contains all complete isomorphisms of $G_p$ within $G_d$. This divide-and-conquer strategy is simple yet effective: reducing the matching task to motif-level computations achieves computational efficiency through parallelization, and reusing motif embeddings avoids the redundant searches common in tree-search-and-prune approaches.

To make this process concrete, we now present a step-by-step example of finding all subgraph isomorphisms. As a first step, we select a set of motifs to be used to decompose $G_p$. Because the number of unique motifs grows rapidly with graph size, we limit ourselves to a small, predefined subset. The motif set $\mathcal{M}$ can be predefined or adaptively selected based on the characteristics of the data graph. The efficiency of the overall decomposition process depends heavily on this choice. We discuss detailed motif selection strategies in Section~\ref{results:motif_selection}, and for now, we focus on demonstrating the workflow.

In this example, we only use two motifs: $M_2$, a simple 2-node edge, and $M_4$. The embeddings of $M_2$, $Res(M_2)$, are always available since they can be obtained by enumerating all edges in $G_d$, and hence serve as a default primitive in our method. Building on these primitives, $Res(M_4)$ can be computed by calling the complete workflow of our solver, which constructs larger motif embeddings from smaller ones like $Res(M_2)$. We defer the details of this computation to Section~\ref{subsec:motif_sequences}, where the underlying components are formalized in Algorithm~\ref{alg:build-motif-db-recursively}; for now, we assume that both $Res(M_2)$ and $Res(M_4)$ have been computed and stored in tabular form like a dataframe structure, as illustrated in Fig.~\ref{fig:motivating_example}(c). Each row in the dataframe corresponds to a matched subgraph in the data graph, and each column represents the vertex indices assigned to the corresponding positions in the motif.

With the motif embeddings prepared, the next step is to decompose the pattern graph using these motifs as templates. There are many valid ways to decompose a pattern graph - we only need to find one. We prioritize larger motifs, as they tend to reduce the number of decomposition steps, a design choice we validate empirically in Sec.~\ref{results:motif_selection}. Additionally, we require that adjacent components share overlapping vertices, which serve as join key constraints during the \textit{join} phase, as detailed later in this section and formalized in Algorithm~\ref{alg:join-filter}. 

In Fig.~\ref{fig:motivating_example}(d), we illustrate two valid decompositions using only $M_4$, with each slice $S_i$ highlighted in a different color. For this example, we choose one such decomposition: $S_0: v_2 - v_3 - v_4 - v_5$ and $S_1: v_5 - v_0 - v_1 - v_2$. Here, $S_i$ denotes the $i$-th graph slice of the decomposition, and the two slices share overlapping vertices at $v_2$ and $v_5$. To clarify the terminology, $M_4$ represents the generic motif structure used as a reusable matching template, while each $S_i$ is a concrete subgraph of $G_p$ with fixed vertex assignments. Since we use $M_4$ for decomposition, both $S_0$ and $S_1$ are instances of $Res(M_4)$, and together they reconstruct the full cycle of $G_p$.

Our decomposition records the overlapping vertices between slices, which later serve as constraints in the \textit{join} phase. In this example, the overlaps occur at $v_2$ and $v_5$, corresponding to the constraints $S_0[0] = S_1[3]$ and $S_0[3] = S_1[0]$. We now proceed to join the motif embeddings corresponding to the slices. For clarity, we denote the two dataframes as $Res(M_4)_\text{L}$ and $Res(M_4)_\text{R}$, representing the left and right sides of the join operation. Based on the overlapping nodes identified during decomposition, the join constraints are defined as
$Res(M_4)_\text{L}[0] == Res(M_4)_\text{R}[3]$ and
$Res(M_4)_\text{L}[3] == Res(M_4)_\text{R}[0]$,
where the bracket notation refers to entire columns of vertex assignments at the respective motif embeddings. These constraints are used as join keys, implemented as hash joins in standard dataframe libraries and relational database systems, whose cost grows linearly with the input and output table sizes. The \textit{join} step eliminates invalid combinations and retains only those pairs of embeddings that satisfy the required boundary conditions. Fig.~\ref{fig:motivating_example}(e) illustrates the join operation and its output to clarify the overall process.

\begin{algorithm}[b]
\footnotesize
\caption{$\Delta$-Motif Join-and-Filter Loop: $R$ collects the final result, $D$ is the decomposition of the pattern graph into motifs}
\label{alg:join-filter}

\begin{algorithmic}[1]
\Function{$\Delta$-Motif}{$G_p, DB$}
    \State $R \leftarrow \emptyset$
    \State $D \leftarrow \Call{decompose}{G_p, DB}$
    \State $M, C \leftarrow \Call{NextMotifToProcess}{D}$ \Comment{$\langle \mathcal{M}, Constraints \rangle$}
    
    \While{$M \neq \emptyset$}
        \State $R \leftarrow \Call{InnerJoin}{R, M, C}$ \Comment{Join}
        \State $R \leftarrow \Call{FilterOverlappingNodes}{R}$ \Comment{Filter}
        \State $M, C \leftarrow \Call{NextMotifToProcess}{D}$    
    \EndWhile
    
    \State \Return $R$
\EndFunction

\end{algorithmic}
\end{algorithm}

Although the \textit{join} operation enforces the specified boundary constraints between motif embeddings, it may still produce combinations with unintended vertex overlaps. These arise when two embeddings share additional vertices beyond the expected overlaps, violating the one-to-one vertex mapping required for valid subgraph isomorphism. To address this, we introduce a \textit{filter} step that removes such invalid combinations. Fig.~\ref{fig:motivating_example}(f) illustrates an example of this issue. In general, the number of duplicated vertices in each joined row should exactly match the number of join constraints. Any row containing additional duplicates is deemed invalid and discarded. This is similar to the pruning phase of a tree-search. 

By applying this \textit{join}-and-\textit{filter} process, we ensure that only valid subgraph isomorphisms are retained in the final result. \textit{Algorithm~\ref{alg:join-filter}} outlines the procedure at a high-level. In the following subsections, we formalize this methodology and show how it enables a highly parallelizable, data-centric algorithm built entirely on open-source, commodity data science tools, without requiring any custom GPU kernels. Nevertheless, the approach delivers strong performance and scalability.

\subsection{Pattern Graph Decomposition Using Motifs}
We now generalize the decomposition and show that any pattern graph can be decomposed using a broader set of motifs.

In principle, many motifs could be used for decomposing the graph; we typically restrict ourselves to a small, carefully-chosen family that balances coverage and computational cost. The choice of motifs has a significant impact on the performance of the algorithm. Larger motifs do not necessarily guarantee faster execution, as their embeddings $Res(M)$ often result in more expensive join operations. Including more motifs in the set also increases the upfront preparation time. In section \ref{results:motif_selection}, we discuss how to select an appropriate motif set based on the characteristics of both the data and pattern graphs, and support these design choices with empirical results. 

Given a defined set of motifs $\mathcal{M}$, we now turn to decomposing the pattern graph $G_p$ into smaller slices. At each iteration, we traverse the chosen motif set in descending order of size, favoring larger motifs to reduce the number of decomposition steps. For each motif in the set, we invoke a subgraph isomorphism solver, such as VF2, to find a valid matching slice $S_i$. If a match is found, it becomes the next slice in the decomposition; otherwise, we continue to the next motif in the list. VF2 is chosen here because it quickly returns a single match, which suits our needs for this iterative, one-slice-at-a-time decomposition process. Because this search runs on the small pattern graph and stops at the first match, its cost depends only on the pattern size, not on the much larger data graph.

While this may seem counterintuitive as we replace one subgraph isomorphism problem with another, we highlight the following: 1) instead of performing subgraph isomorphism on the data graph, we apply it to the typically much smaller pattern graph, 2) the motifs we seek are typically orders of magnitude smaller than both the data and pattern graph, and 3) in our motif decomposition, we only require finding a single match of the motif in the pattern, allows us to terminate the search early rather than exhaustively enumerating all possible matches. Under these conditions, the tree-search based algorithm is actually more suitable and efficient for this task. Lastly, the motif decomposition searches for {\bf a} decomposition rather than the ideal decomposition.

Once a slice $S_i$ is identified, the algorithm inspects all vertices in $S_i$ to determine which have outgoing edges that remain uncovered in the current pattern graph. These vertices are designated as boundary nodes. Boundary nodes serve as connection points for subsequent slices. For example, in Section \ref{subsec:motif_method}, vertices $v_2$ and $v_5$ from $S_0$ play this role. The non-boundary nodes from $S_i$ are removed from the pattern graph to form a reduced graph for the next iteration, while the boundary nodes are retained to ensure overlap with subsequent matches. This procedure is repeated in the following iterations. Each new slice is matched using the reduced pattern graph, and its overlapping vertices with the union of previously selected slices are recorded. These overlapping vertices define structural dependencies across slices and are stored as join constraints. These constraints are later used in the \textit{join} step, formalized in Algorithm~\ref{alg:join-filter}, to reassemble the full pattern structure.

As a result of the decomposition process, we obtain a sequence of slices $S_0, S_1, \dots, S_n$ along with their corresponding list of motifs. This sequence determines the order in which the motifs appear within the pattern graph, and consequently, the order in which the associated $Res(M)$ dataframes are joined during reconstruction.

In practice, the decomposition step is computationally inexpensive and contributes only a negligible portion to the overall runtime, though motif selection, ordering, and decomposition strategy remain promising avenues for future optimization.

\subsection{From a Single Motif to Computing Embeddings of Motif Sequences}
\label{subsec:motif_sequences}
In the above, we decomposed the pattern graphs into a sequence of motifs. In this subsection, we show how we can extract the embeddings of those motifs from the data graph and how we connect the results of multiple motif searches.
As a simple case, searching for motif $M_2$ is equivalent to returning all edges of $G_d$ excluding self-loops $a-a$. Formally, $Res(M_2) = E_d \setminus E_{self}$, where $E_{self}$ is the set of self-loops.

To compute embeddings of larger motifs, we apply a sequence of \textit{join}-and-\textit{filter} operations using an algorithm described in Algorithm 1. For example, consider the motif $M_3$, defined as a simple path: $a -b -c$ where $a \ne b \ne c$. Its embeddings, denoted $Res(M_3)$, can be computed by inner-joining two instances of $Res(M_2)$ on the shared node $b$. This can be written as: $Res(M_3) = \text{L} \bowtie_{\text{L}[1] == \text{R}[0]} \text{R}$, where both $\text{L}$ and $\text{R}$ refer to the same table $Res(M_2)$. The \textit{join} captures all three-node paths, and is followed by a filtering step to eliminate invalid embeddings. For example, since both edges $a - b$ and $b - a$ exist in the undirected data graph, the resulting table may incorrectly include the cyclic pattern $a - b - a$. The \textit{filter} removes such entries and ensures that all three vertices in each triplet are distinct. This entire process can be carried out in a single call to our motif solver. The resulting set $Res(M_3)$ is stored in a dataframe format.

\begin{algorithm}[b]
\footnotesize
\caption{Build Motif Database Recursively}
\label{alg:build-motif-db-recursively}

\begin{algorithmic}[1]
\Function{BuildDatabase}{$G_d, \mathcal{M}$}
    \State $DB \leftarrow \{(u,v) : (u,v) \in E_d\}$ 
    \Comment{DB is $Res(M_{2})$}
    \ForAll{$M_i \in \mathcal{M}$}
        \State $Res(M_i) \leftarrow \Call{Delta-Motif}{M_i, DB}$
        \State $DB \leftarrow DB \cup Res(M_i)$
    \EndFor
    \State \Return $DB$
\EndFunction
\end{algorithmic}
\end{algorithm}

Similarly, embeddings of larger motifs can be computed recursively. For example, $Res(M_4)$ can be built by joining $Res(M_3)$ with $Res(M_2)$ on the last and first vertex: $Res(M_4) = \text{L} \bowtie_{\text{L}[2] == \text{R}[0]} \text{R}$, where $\text{L} = Res(M_3)$ and $\text{R} = Res(M_2)$, followed by filtering. This pattern generalizes to arbitrary motif sizes (e.g. $Res(M_5)$ from $(M_3, M_3)$ or $(M_4, M_2)$), as shown in \textit{Algorithm ~\ref{alg:build-motif-db-recursively}}.

\subsection{Performance Increase from Larger Motifs}
The use of larger motifs for decomposing the pattern graph has several benefits: 1) fewer motifs are needed for the decomposition, hence fewer \textit{join}-and-\textit{filter} , which can be substantially fewer than the branching levels explored in a VF2 tree search, 2) larger motifs can increase parallelism in the data pre-processing phase, and 3) larger motifs enable early-stage pruning of invalid solutions, which also takes place during the pre-processing phase.


\section{Analog To Tree Search And Prune}
\label{sec:search_and_prune}
Having presented our join-based algorithm for subgraph isomorphism, we now draw a parallel to traditional tree-search-and-prune approaches. Although not obvious at first, these algorithms share many features while trading off others.

\subsection{Parallelism Vs. Storage}
Most tree-search-and-prune solutions rely on recursion. Some parallelism is possible---typically by enumerating an ordered set of vertices for the first few levels of the tree (e.g., one to three) and parallelizing across them, i.e., an initial BFS to generate starting conditions followed by DFS to find the remaining pattern. Extending parallelism beyond these levels raises several challenges: workload imbalance, a larger memory footprint for storing starting points (similar to generating $Res(M_3)$), and, most critically, synchronization overhead as threads coordinate access to shared data structures. These limits on achievable parallelism also explain why some algorithms handle only patterns with fewer than 10 vertices and edges.

In contrast, most phases of $\Delta$-\textit{Motif} are inherently parallel: each row in our dataframe is an independent candidate, and the same operations apply uniformly across all rows, letting thousands of candidate mappings be processed simultaneously, similar to BFS. This parallelism comes at a cost: our new algorithm has an increased memory footprint. The increased memory requirements stem from two main factors: 1) holding intermediate data generated during join operations before key constraints are applied to prune invalid matches and 2) storing complete tables of intermediate solutions after each join-and-filter iteration.

\subsection{$M_{2}$ and Its Analogy to Parallel Tree-Search}
\label{sec:M2_parallel}
Consider the case where the pattern graph is decomposed only with the $M_2$ motif. In this setup, the number of steps in $\Delta$-\textit{Motif} matches the branching levels explored in a VF2 tree search: the first $M_2$ expansion enumerates all edges, the next expands them into all possible 3-hop paths, then into 4-hop paths, and so on.The key difference is that VF2 moves forward one branch at a time, while $\Delta$-\textit{Motif} processes all branches at a given depth in parallel (similar to a BFS traversal) through joins and filters that run simultaneously across all candidates.

However, the $M_2$-only case shares a limitation with VF2: the same path may be traversed multiple times from different starting points, incurring redundant work. In contrast, using larger motif choices allows $\Delta$-\textit{Motif} to operate on larger structural units, avoiding repeated traversal; invalid candidates are discarded during filtering rather than through recursive backtracking. 

\subsection{$\Delta$-\textit{Motif} and $\Delta$-\textit{Stepping}}
We now return to the analogy with the $\Delta$-\textit{Stepping} algorithm for solving the SSSP problem. Although SSSP and subgraph isomorphism address different problems, both approaches expose additional parallelism by allowing extra work during execution. In $\Delta$-\textit{Stepping}, the parameter $\Delta$ trades computational efficiency for parallelism by allowing vertices to be processed earlier at the cost of potentially revisiting them. Similarly, our $\Delta$-\textit{Motif} exposes parallelism at multiple granularities: across multiple paths all within the same motif, across the entire \textit{join}-and-\textit{filter} operations, and across the preparation of all the motif embeddings in pre-processing computation. These parallelization opportunities apply uniformly to both CPU and GPU backends; we quantify their performance impact in Sec.~\ref{sec:benchmarks}.

\section{Experimental Setups}
\label{sec:problem_setup}
\subsection{Software and Hardware Configurations}
\paragraph*{\bf $\Delta$-Motif Implementation}\
Our CPU implementation leverages both \texttt{Pandas} 2.2.3~\cite{Pandas} and \texttt{NumPy} 2.0.0~\cite{NumPy}, representing motif embeddings and candidate isomorphisms as DataFrames. On GPU, $\Delta$-\textit{Motif} uses \texttt{cuPy}~\cite{Cupy2017} and \texttt{cuDF} 25.6.0~\cite{cudf-25.6.0}, which serve as drop-in replacements for \texttt{NumPy} and \texttt{Pandas}, respectively. Thanks to strong interoperability across the NVIDIA RAPIDS ecosystem and adherence to Apache Arrow standards, the algorithm transitions between CPU and GPU execution with only minimal changes to Python imports. Although subgraph isomorphism is a graph problem, our implementation relies not on \texttt{cuGraph} or \texttt{NetworkX}~\cite{NetworkX} but directly on tabular representations using \texttt{cuDF} and \texttt{cuPy}.

\paragraph*{\bf VF2 and rustworkx}\
We use the VF2 algorithm as our baseline, representing the broader class of tree-search-and-prune methods for subgraph isomorphism. We selected \texttt{rustworkx}, a high-performance Rust implementation that delivers up to 100× speedup than python \texttt{NetworkX} and is used in production systems such as IBM's Qiskit quantum computing framework \cite{qiskit2024}. This ensures that our comparisons reflect true algorithmic differences rather than implementation bottlenecks. 

\paragraph*{\bf GSI}\
We also include GPU-friendly Subgraph Isomorphism (GSI) \cite{GSI2020, gsi_github} as a representative GPU-based baseline designed to improve parallel efficiency through GPU-centric execution strategies. We compare against GSI where the reference implementation is applicable; however, for most of our benchmark configurations involving larger pattern graphs, GSI did not complete within reasonable execution time or encountered out-of-memory errors.

\paragraph*{\bf Hardware Configurations}\
Our GPU-based experiments were conducted on the NVIDIA H200 GPU, part of the Hopper architecture family, equipped with 16,896 scalar processors and 144GB of HBM3e memory. For CPU-based benchmarking and the VF2 baseline, we used an AMD Ryzen Threadripper PRO 3995WX running at 4.3 GHz with 64 physical cores, 128 threads, and 1TB of DDR4-3200 system memory.


\subsection{Benchmark Metrics on Runtime}
The execution time of the $\Delta$-\textit{Motif} algorithm comprises two phases: data preparation and computation. In the data preparation phase, the embeddings of required motif set, $Res(M)$, are computed and stored as dataframes for reuse. The computation phase then decomposes the pattern graph and executes dataframe \textit{join}-and-\textit{filter} operations to produce the final matches.

In many scenarios, the same data graph is queried repeatedly with different pattern graphs, as in quantum circuit compilation~\cite{Nation2023} or streaming subgraph isomorphism~\cite{Duong2021} across social network analysis, chemistry, and biology. Here the data preparation phase can be performed once, cached, and reused, excluding its cost from subsequent queries and leaving only the pattern-searching phase, delivering significant runtime savings and further speedups. To highlight this potential, we also report results with data preparation time omitted, in Section~\ref{sec:benchmarks} marked as $\Delta$-\textit{Motif} GPU$^*$.

We observed minimal runtime variance across repeated runs using VF2 and $\Delta$-\textit{Motif}; therefore, we report results from a single execution for each $(G_d, G_p)$ configuration.

\section{Benchmarks and Results}
\label{sec:benchmarks}
            

\begin{table*}[!t]
\footnotesize
\centering
\caption{Observed speedup ranges of $\Delta$-\textit{Motif} (GPU) over VF2 across pattern sizes and data graph configurations. For each configuration, we report the mean, minimum, and maximum speedups measured over all random seeds. GSI performance is discussed in the main text but omitted from this table, as results were unavailable for most problem sizes.}
\label{tab:speedup_range}
\setlength{\tabcolsep}{4pt}
\renewcommand{\arraystretch}{1.2}
\begin{tabular}{
    c c
    | r r r
    | r r r
    | r r r
    | r r r
    | r r r
}
\toprule
\multicolumn{2}{c|}{\textbf{Data Graph}} &
\multicolumn{15}{c}{\textbf{Pattern Graph Size}} \\
\cmidrule(lr){3-17}
\multirow{2}{*}{\textbf{Type}} &
\multirow{2}{*}{\textbf{Size}} &
\multicolumn{3}{c|}{\textbf{20}} &
\multicolumn{3}{c|}{\textbf{40}} &
\multicolumn{3}{c|}{\textbf{60}} &
\multicolumn{3}{c|}{\textbf{80}} &
\multicolumn{3}{c}{\textbf{100}} \\
\cmidrule(lr){3-5} \cmidrule(lr){6-8} \cmidrule(lr){9-11} 
\cmidrule(lr){12-14} \cmidrule(lr){15-17}
& &
\textbf{min} & \textbf{mean} & \textbf{max} &
\textbf{min} & \textbf{mean} & \textbf{max} &
\textbf{min} & \textbf{mean} & \textbf{max} &
\textbf{min} & \textbf{mean} & \textbf{max} &
\textbf{min} & \textbf{mean} & \textbf{max} \\
\midrule
\multirow{2}{*}{\makecell{Heavy\\Hex}}
  & 1990 & 0.64 & 2.96 & 7.84 & 1.61 & 12.78 & 151.11 & 1.85 & 14.05 & 210.91 & 1.83 & 15.62 & 179.33 & 2.48 & 13.03 & 160.35 \\
  & 4485 & 2.76 & 10.09 & 22.58 & 5.70 & 35.76 & 238.03 & 6.84 & 42.43 & 294.80 & 9.95 & 49.53 & 269.72 & 9.18 & 42.36 & 415.20 \\
\midrule
\multirow{2}{*}{\makecell{Square\\Grid}}
  & 1600 & 0.91 & 3.29 & 46.98 & 1.49 & 6.24 & 78.85 & 2.21 & 11.58 & 313.95 & 2.44 & 12.65 & 135.46 & 1.57 & 15.20 & 99.58 \\
  & 3600 & 2.75 & 7.79 & 127.64 & 5.17 & 17.50 & 146.19 & 7.25 & 30.11 & \textbf{594.92} & 8.32 & 36.50 & 306.00 & 8.15 & 53.43 & 269.03 \\
\bottomrule
\end{tabular}
\end{table*}

We evaluate our GPU-accelerated, motif-based subgraph isomorphism approach on graph workloads motivated by quantum circuit compilation, which feature structured data graphs and comparatively large pattern graphs.

\subsection{Subgraph Enumeration With Synthetic Subgraphs}
In this section, we benchmark $\Delta$-\textit{Motif} on quantum device topologies relevant to layout selection in circuit compilation. We use the heavy-hex lattice and the 2D square grid topologies, shown in Fig.~\ref{fig:motifs}(a) and (b). For each topology, we consider two sizes that reflect the projected number of qubits for near-future quantum devices: 1990 and 4485 vertices for the heavy-hex topology and 1600 and 3600 vertices for the square grid topology. Thus, our data graphs represent real-world graphs. For $\Delta$-\textit{Motif}, we select motif sets tailored to each topology: \{$M_2$, $M_4$\} for heavy-hex and \{$M_2$, $M_{4\text{-}O}$, $M_{6\text{-}O}$\} for square grid.   

For each data graph, we generate random connected subgraphs as pattern graphs of sizes ${20, 40, 60, 80, 100}$ vertices. For each pattern size, we sample 200 random seeds to capture structural variability. For each case, we enumerate all isomorphic mappings using $\Delta$-\textit{Motif} on GPU and the VF2 baseline on CPU. Speedups are reported as the ratio $T_{\text{VF2}} / T_{\Delta\text{-}Motif}$, with minimum, mean, and maximum values shown in Table~\ref{tab:speedup_range}.

Table~\ref{tab:speedup_range} confirms a consistent advantage for $\Delta$-\textit{Motif} GPU across all configurations, with mean speedups ranging from $3\times$ on small patterns (which typically under-utilize the GPU) and up to $53.43\times$ on the largest cases. For smaller patterns, $\Delta$-\textit{Motif} under-utilizes the GPU, and fixed launch and data-movement overheads constitute a larger fraction of the total runtime. Larger data graphs generally yield higher speedups; for example, increasing the square grid size from 1600 to 3600 vertices raises the mean speedup for 100-node patterns from $15.20\times$ to $53.43\times$. Medium-to-large patterns (60 nodes and above) generally achieve the highest speedups, particularly for the 3600-node square grid and 4485-node heavy-hex, where the parallel \textit{join}-and-\textit{filter} phase is fully leveraged, reaching maximum observed speedups of $415.20\times$ and $594.92\times$, respectively. While variability is observed across seeds, the minimum speedup remains above the VF2 baseline in nearly all cases. The few low-minimum cases (e.g., heavy-hex with 20-node patterns) arise when the pattern size is sufficiently small that GPU launch overhead dominates the runtime.

We additionally compare $\Delta$-\textit{Motif} against GSI where results are available. On smaller instances, such as the 1600-node square grid with 20-node patterns, the two methods show comparable average performance, measured by the mean speedup $T_{\text{GSI}} / T_{\Delta\text{-}Motif}$, with maximum speedups of up to $17\times$. As problem difficulty increases, the performance gap widens. For the 3600-node square grid with 20-node patterns, $\Delta$-\textit{Motif} is on average $2\times$ faster than GSI, with maximum speedups reaching $63\times$. For larger patterns, GSI fails to complete within a reasonable time window, while $\Delta$-\textit{Motif} continues to scale effectively.

Overall, the results show that $\Delta$-\textit{Motif} consistently outperforms VF2 and increasingly outperforms GSI as problem size and complexity grow, with the largest benefits realized on large data graphs and medium-to-large pattern sizes that maximize GPU utilization. We also note that pattern graph decomposition accounts for less than 1\% of the total runtime in all cases, with end-to-end execution completing in just a few seconds.

\subsection{Performance Impact of Motif Selection}
\label{results:motif_selection}
\begin{figure}[!b]
    \centering
    \includegraphics[width=0.42\textwidth]{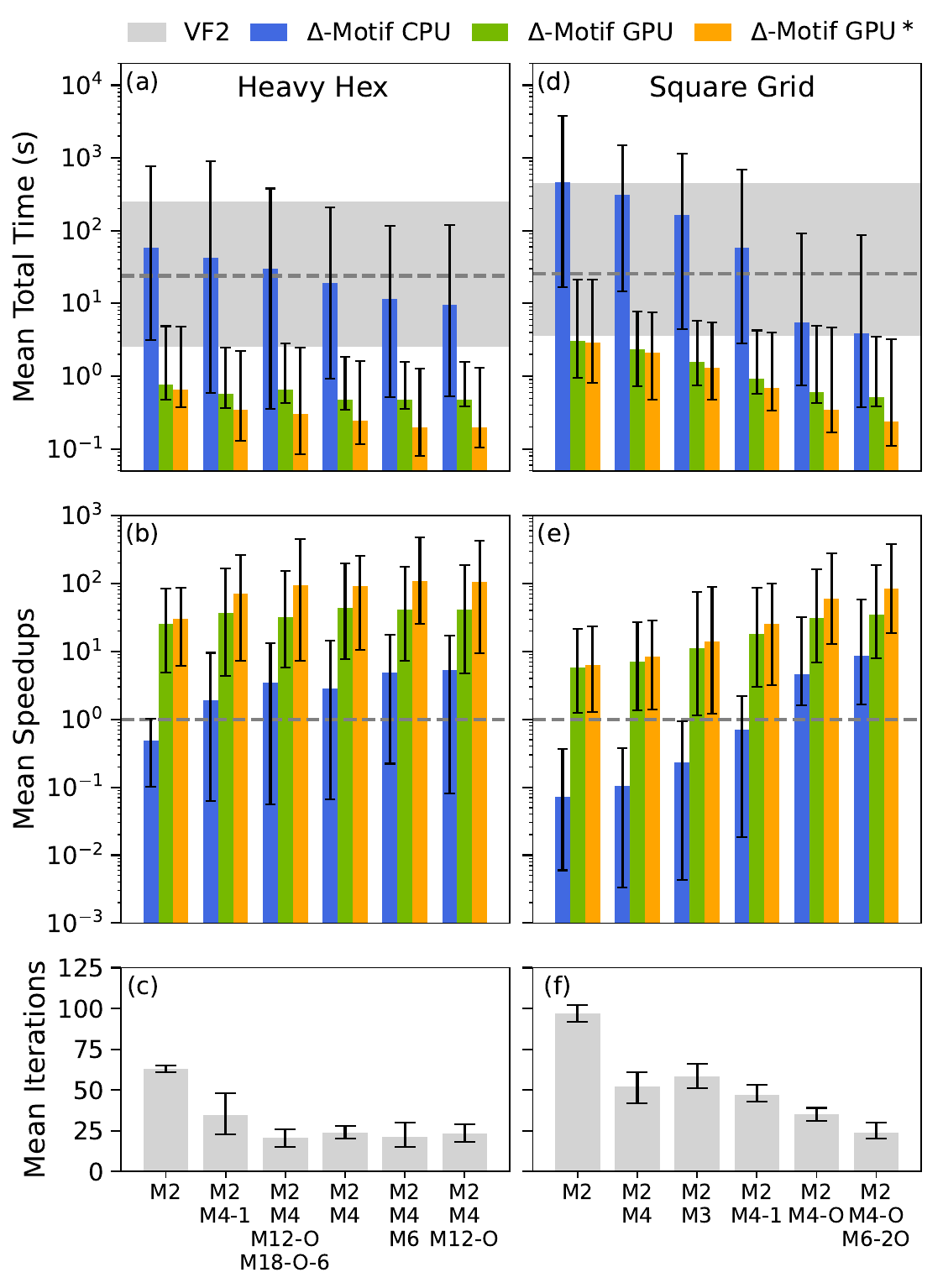}
    \caption{
    Performance comparison on two data graph topologies: (a)–(c) heavy-hex lattice (4,485 nodes) and (d)–(f) 2D square grid (3,600 nodes), using 50 random 60-node pattern graphs. 
    Panels show (a)(d) total execution time, (b)(e) speedup relative to VF2, and (c)(f) number of decomposition iterations. 
    Bars represent mean values, with error bars indicating the minimum and maximum. 
    Results are reported for VF2 and $\Delta$-\textit{Motif} on CPU and GPU, as well as $\Delta$-\textit{Motif} GPU$^*$ with data preparation excluded. GSI did not complete when the pattern graph had 60 vertices.
}

    \label{fig:motif_selection_rotated}
\end{figure}
The experiments in Fig.~\ref{fig:motif_selection_rotated} assess how different motif selection strategies influence runtime, speedup, and decomposition behavior on two data graphs. Each experiment uses 60-node random pattern graphs and compares VF2 with $\Delta$-\textit{Motif} executed on CPU, on GPU, and on GPU with data preparation time excluded (marked as GPU$^*$).\footnote{We focus on 60-node patterns rather than 100-node ones, since CPU execution times for the latter were too long.}. Bars show mean values over 50 random seeds, with error bars representing the minimum and maximum values. The three subplot columns present total runtime, speedups over VF2, and the average number of  iterations for pattern decomposition using the chosen motif set.

Across both heavy-hex and square grid topologies, the GPU version of $\Delta$-\textit{Motif} achieves the lowest execution times, often providing speedups of one to two orders of magnitude compared to VF2. Excluding the one-time data preparation step yields even greater gains, highlighting the benefit of caching in repeated-query scenarios. The CPU version also outperforms VF2 in most cases, although the margins are smaller.

The choice of motif strongly affects performance. The baseline motif $M_2$ (Sec.~\ref{sec:M2_parallel}), functionally similar to a parallelized tree search, delivers only modest speedups because it produces as many decomposition iterations as the pattern graph has edges. Larger motifs substantially reduce the iteration count, lowering the volume of intermediate data processed during the \textit{join}-and-\textit{filter} stages and often translating directly into faster runtimes. However, this benefit depends on the topology of the data graph. In higher connectivity graphs such as the square grid, larger motifs tend to create much larger intermediate joins, which increase memory requirements and can offset iteration savings. In contrast, on lower-connectivity heavy-hex graphs, the same motifs are more effective, benefiting from iteration reduction without incurring excessive intermediate results. Lastly, in Fig.~\ref{fig:motif_selection_rotated}(c) and (f), the average number of $M_2$ iterations matches the average edge count of the pattern graphs and shows little variance, whereas decompositions with larger motifs exhibit much greater variability.

Importantly, simply increasing motif size does not guarantee better speedups. The most effective motifs also appear frequently as repetitive patterns in the data graph. For example, $M_{12\text{-}O}$ and $M_{18\text{-}O\text{-}6}$ for heavy-hex, and $M_{4\text{-}O}$ and $M_{6\text{-}2O}$ for square grids, outperform other motifs of similar size with less representative structure. These results indicate that motif selection should be topology-aware: low-connectivity graphs benefit from large, frequent motifs to reduce iterations, while high-connectivity graphs often require more moderate motifs to balance iteration count with manageable join complexity, allowing the GPU-accelerated \textit{join}-and-\textit{filter} pipeline to operate at peak efficiency.

\subsection{Impact of Solution Size on Performance}
\label{results:solution_size}

\begin{figure}[!t]
    \centering    \includegraphics[width=0.48\textwidth]{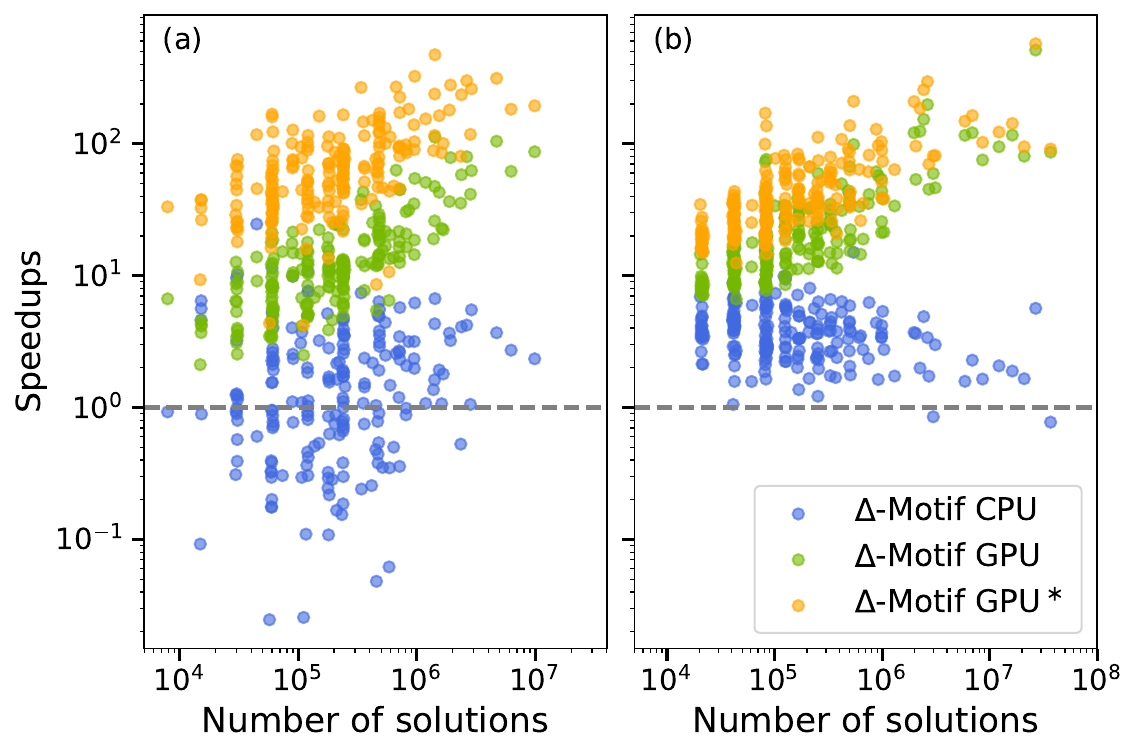}
    \caption{
    Speedup of $\Delta$-\textit{Motif} over VF2 for two data graph configurations: (a) heavy-hex lattice with 4,485 nodes and (b) 2D square grid with 3,600 nodes. Results are shown for $\Delta$-\textit{Motif} on CPU, GPU, and GPU with data preparation excluded (denoted GPU$^*$). The x-axis indicates the number of isomorphic solutions, and the y-axis shows the speedup relative to VF2. Each point corresponds to a distinct 60-node random pattern graph generated from one of 200 random seeds.
    }
    \label{fig:speedup_vs_solution}
\end{figure}

Fig.~\ref{fig:speedup_vs_solution} investigates how the number of outputted solutions, aka the number of enumerated isomorphic subgraphs, impacts the relative performance of $\Delta$-\textit{Motif} compared to VF2. For this plot, we fixed the pattern graph size at 60 nodes but sampled 200 random seeds to generate distinct pattern subgraphs. These variations lead to different numbers of isomorphic matches, which directly impact the size of intermediate data processed during computation. For this experiment, we used motif sets \{$M_2$, $M_{4\text{-}1}$\} on the heavy-hex lattice and \{$M_2$, $M_{4\text{-}O}$\} on the square grid.

Across both heavy-hex and square grid data graphs, GPU $\Delta$-\textit{Motif} implementation consistently outperforms both its CPU counterpart and VF2. Speedups increase with the number of solutions, as larger solution sets generate more intermediate data and expose greater parallelism for the GPU to exploit. Square grid graphs, with their higher connectivity and branching possibilities, also present a more challenging search space and larger intermediate data volumes than heavy-hex graphs. When the number of solutions is small, the CPU can remain relatively efficient because intermediate data volumes are still moderate and GPU parallelism is underutilized. Comparing GPU to GPU$^*$ reveals that data preparation can be costly in some cases. This gap is more evident in cases with smaller solution sets, where compute time is short and preparation overhead dominates. As the solution count grows, speedups for GPU$^*$ version can exceed two orders of magnitude, showing the efficiency of parallel \textit{join}-and-\textit{filter} operations on substantial datasets.

On CPU, $\Delta$-\textit{Motif} outperforms VF2 in about half of the heavy-hex cases and in most square-grid cases (Fig.~\ref{fig:speedup_vs_solution}), confirming that the tabular formulation is also effective without GPU acceleration. Overall, $\Delta$-\textit{Motif} benefits most from GPU acceleration at scale, and reusing precomputed data in repeated-query scenarios can significantly amplify these gains.

\subsection{End-to-End Transpilation Benchmark}
\label{results:transpilation}
As described in Sec.~\ref{subsec:quantum_compilation}, layout selection comprises layout generation (subgraph isomorphism enumeration) and layout scoring (ranking candidates by estimated execution fidelity based on device-specific qubit quality metrics).
Note that layout scoring is also critical to end-to-end performance, as an inefficient implementation can significantly impact runtime. In $\Delta$-\textit{Motif}, we chose to compute the scores at the level of motif embeddings $Res(M)$ rather than waiting for full solutions. The fidelity scores are stored as a new column in $Res(M)$ and updated incrementally after each subsequent \textit{join} by multiplying the joined fidelity columns. By the end of the process, the total fidelity score for each subgraph is directly available. This scoring approach benefits from our tabular data structure, allowing fidelity evaluations to be embedded directly into the same relational operations used for candidate generation. As a result, our scoring extension adds trivial overhead and remains highly efficient, as shown in our later benchmarks.

In this benchmark, we integrated the $\Delta$-\textit{Motif} algorithm into a complete quantum transpilation pipeline for layout selection. We compared our approach against Qiskit’s \texttt{VF2PostLayout} transpiler pass \cite{qiskit2024}, which relies on \texttt{rustworkx}’s VF2 implementation for layout generation, followed by a sequential Python-based scoring function for candidate scoring \cite{Nation2023}.\footnote{The scoring stage is sequential for undirected data graphs, which we use because they yield more isomorphisms. While some quantum hardware supports only directed two-qubit operations, reversing the direction in the postprocessing is trivial in practice, allowing us to consider a much larger space of feasible layouts using the undirected model.} Note that although the design of scoring cost functions is an active research topic in quantum computing, it is not the focus of this work. Here, we adopt the same cost function as used in \cite{Nation2023}.

The data graph was derived from the qubit connectivity of IBM’s \texttt{ibmq\_fez}, a state-of-the-art 156-qubit quantum processor (represented as a 156-node graph). We benchmark our pipeline on a diverse set of pattern graphs obtained from quantum circuits in the \texttt{QASMBench} benchmark suite~\cite{qasmbench}, covering sizes from 30 to 70 qubits and including a scaling set of Bernstein–Vazirani (BV) circuits.\footnote{Before layout selection, all input logical circuits are pre-transpiled with Qiskit (opt\_level=1) to ensure that every two-qubit operation is mapped to physically connected qubits, making the resulting pattern graphs valid subgraphs of the device connectivity.} 

\begin{figure}[t]
    \centering
    \includegraphics[width=0.48\textwidth]{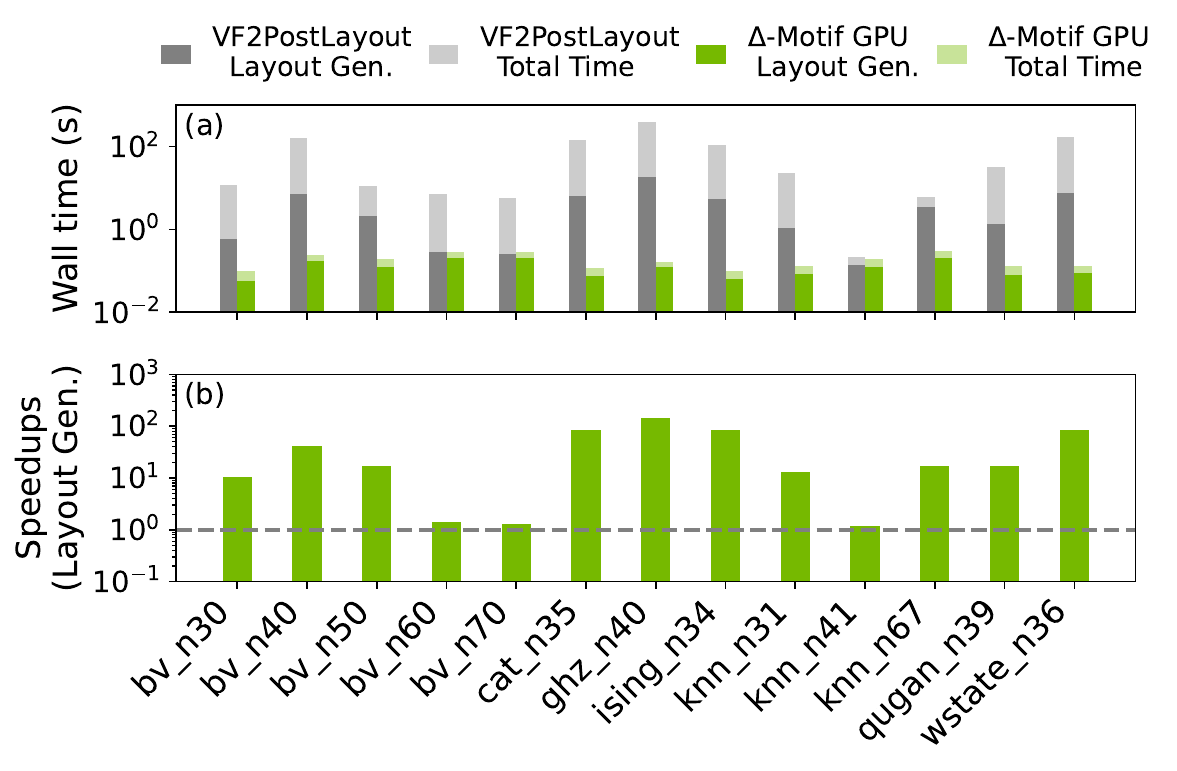}
    \caption{
    Quantum layout selection benchmarks on IBM’s \texttt{ibmq\_fez} topology (156 qubits), comparing $\Delta$-\textit{Motif} with Qiskit’s \texttt{VF2PostLayout} across selected circuits from the \texttt{QASMBench} suite. 
    (a) Wall-clock time for completing layout selection, broken into layout generation and other components (primarily layout scoring). In some VF2 cases, scoring dominates the runtime. 
    (b) Speedups in layout generation time, comparing $\Delta$-\textit{Motif} GPU against VF2. 
    Labels follow the format \texttt{<algorithm>\_n<qubits>}, e.g., \texttt{bv\_n30} denotes a 30-qubit Bernstein–Vazirani circuit.
}
    \label{fig:transpilation}
\end{figure}

Results are presented in Fig.~\ref{fig:transpilation}. Panel (a) reports the total wall-clock time for layout selection, broken down into layout generation and other components (primarily scoring). For \texttt{VF2PostLayout}, the sequential Python-based scoring often dominates the runtime, whereas $\Delta$-\textit{Motif} maintains highly efficient scoring through its tabular design. To highlight our core algorithmic innovation, panel (b) only reports speedups for layout generation (i.e., subgraph isomorphism) relative to VF2. Across this varied workload, GPU implementations of $\Delta$-\textit{Motif} consistently outperform VF2 by up to two orders of magnitude. For very small circuits or cases with small number of layouts, all methods complete layout generation quickly and GPU launch overheads dominate the runtime, so speedups are modest. However in these cases, the sequential layout scoring in VF2 becomes the main bottleneck. In contrast, when circuits yield many valid layouts on the 156-qubit device, the larger intermediate search spaces amplify GPU advantages. This trend is especially clear in the BV scaling experiments, where the 40-qubit BV circuit produces the most layouts and exposes greater parallelism.  

These results illustrate a realistic challenge in quantum layout selection: even when data graphs are only a few hundred nodes, large pattern graphs can yield enormous search spaces. Here, \texttt{VF2PostLayout} suffers from sequential bottlenecks in both layout generation and scoring, whereas $\Delta$-\textit{Motif} delivers strong and stable execution with total runtimes well under one second. Our tabular-based scoring adds minimal overhead. As quantum devices scale to hundreds or even thousands of qubits in the coming years, the scalability trends observed in this work provide strong evidence that our method can continue to deliver substantial performance benefits.

\section{Conclusion}
\label{sec:conclusion}
In this paper, we introduced $\Delta$-\textit{Motif}, a parallel subgraph isomorphism algorithm motivated by the layout-selection bottleneck in quantum circuit compilation. It transforms the traditional sequential search into a massively parallel data-processing task, decomposing pattern graphs into smaller motifs and recombining them through database primitives without specialized GPU kernels. 



We implement $\Delta$-\textit{Motif} using open-source, commodity software that exploits the parallelism of NVIDIA GPUs. By avoiding hardware-specific kernels and leveraging optimized RAPIDS libraries, our design remains portable across accelerator platforms. On quantum device topologies, this implementation consistently achieves order-of-magnitude performance improvements over VF2, with observed speedups of up to $595\times$, and outperforms the GPU-based approach GSI where direct comparison is applicable. In end-to-end transpilation benchmarks, $\Delta$-\textit{Motif} delivers up to two orders of magnitude faster layout generation on a state-of-the-art 156-qubit quantum processor. To the best of our knowledge, $\Delta$-\textit{Motif} is the first GPU-based subgraph isomorphism approach implemented entirely using database-style relational primitives. As quantum devices scale to thousands of qubits, the scalability trends shown here position $\Delta$-\textit{Motif} as a practical tool for next-generation quantum compilation and open new research directions at the intersection of graph algorithms, quantum, and high-performance computing.

\section*{Acknowledgment}
We thank Bradley Dice for valuable discussions and support. We are also grateful to our colleagues at Q-CTRL, NVIDIA, and Oxford Quantum Circuits, whose technical expertise, product engineering, and design efforts contributed significantly to the results presented in this paper.

\bibliographystyle{IEEEtran}
\bibliography{reference}

\end{document}